\documentclass{article}

\usepackage{microtype}
\usepackage{graphicx}
\usepackage{subfigure}
\usepackage{booktabs} 

\usepackage{hyperref}

\newif\ifmsaccepted
\msacceptedtrue

\ifmsaccepted
\usepackage[accepted]{icml2024-diffxyz}
\newcommand{\repourl}{\url{https://github.com/cory-b-scott/diff_ifs}}
\else
\usepackage{icml2024-diffxyz}
\newcommand{\repourl}{\url{https://anonymous.4open.science/r/diff_ifs-ADC3} (anonymized URL)}
\fi
\usepackage{amsmath}
\usepackage{amssymb}
\usepackage{mathtools}
\usepackage{amsthm}

\usepackage{graphicx}

\usepackage[capitalize,noabbrev]{cleveref}

\theoremstyle{plain}

\theoremstyle{definition}

\theoremstyle{remark}

\usepackage[textsize=tiny]{todonotes}

\usepackage[
backend=biber,
style=ieee,
citestyle=ieee,
]{biblatex}
\AtBeginBibliography{\small}

\icmltitlerunning{Differentiable IFS Fractals}

\newcommand{\fractalsubfigwidth}{0.18\linewidth}
\newcommand{\examplesubfigwidth}{0.15\linewidth}
\newcommand{\kochsubfigwidth}{0.37\linewidth}

\begin{document}

\twocolumn[
\icmltitle{Differentiable Iterated Function Systems}



\icmlsetsymbol{equal}{*}

\begin{icmlauthorlist}
\icmlauthor{Cory B. Scott}{equal,colo}
\end{icmlauthorlist}

\icmlaffiliation{colo}{Department of Mathematics and Computer Science, Colorado College, Colorado Springs, USA}

\icmlcorrespondingauthor{Cory B. Scott}{cbs@coloradocollege.edu}

\icmlkeywords{Differentiability, IFS, differentiable rendering}

\vskip 0.3in
]



\printAffiliationsAndNotice{}  
\begin{abstract}
This preliminary paper presents initial explorations in rendering Iterated Function System (IFS) fractals using a differentiable rendering pipeline. Differentiable rendering is a recent innovation at the intersection of computer graphics and machine learning. A fractal rendering pipeline composed of differentiable operations opens up many possibilities for generating fractals that meet particular criteria. In this paper I demonstrate this pipeline by generating IFS fractals with fixed points that resemble a given target image - a famous problem known as the \emph{inverse IFS problem}. The main contributions of this work are as follows: 1) I demonstrate (and make code available) this rendering pipeline; 2) I discuss some of the nuances and pitfalls in gradient-descent-based optimization over fractal structures; 3) I discuss best practices to address some of these pitfalls; and finally 4) I discuss directions for further experiments to validate the technique.  
\end{abstract}

\section*{Introduction and Prior Work}
Fractals are ubiquitous objects in computer graphics, mathematical art, and data analysis. A common way to generate fractal images is the \textit{iterated function system} (IFS), defined in detail below. In this paper, I demonstrate a system for drawing IFS fractals written in Pytorch, a programming language with automatic differentiation capability. The advantage of doing this in Pytorch is that it enables us to optimize through the fractal rasterization process, so that the generated IFS fractal resembles a target image. This paper is inspired by the recent work "Differentiable Drawing and Sketching", by Mihai et al. \cite{mihai2021differentiable}. Those authors presented a framework for differentiating through the rasterization of lines and curves, using the insight that the distance from a pixel to a line can act as a differentiable proxy for rasterization. Since distance estimators have a long history of use in the fractal rendering community \cite{christensen2011distance,hart1989ray,mcguire2014numerical}, it is natural to combine some of those ideas with the techniques from Mihai et al. I will first define some of the background needed to understand the problem, and then I will introduce how the rendering pipeline works. Throughout this paper, I will use the Koch curve (a famous fractal \cite{koch1904courbe}) as a test case. 
\subsection*{IFS Fractals}
I follow the definition of IFS fractal given by Barnsley \cite{barnsley1993fractals} and later by Heptig \cite{hepting1991rendering}. Let $F_1, F_2 \ldots F_k$ be a set of affine transformations of the Euclidean plane $\mathbb{R}^2$, and for any region $\mathcal{S} \subset \mathbb{R}^2$ let $F_i(\mathcal{S})$ denote the image of $\mathcal{S}$ under $F_i$.  The \emph{Hutchinson} transform \cite{hutchinson1981fractals} $H(S)$ is the union of the images of $\mathcal{S}$ under all of the $F_i$: $H(\mathcal{S}) = \cup_{i = 1}^n F_i(\mathcal{S}) $.

The \emph{attractor} of an IFS is a limit figure (not necessarily unique) which results from applying this process infinitely many times, i.e. a region $A$ such that $A = H(A)$. For an attractor to exist, our set of affine transformations must on average be \textit{contractive}, meaning that they shrink the distance between points in the plane. 

\begin{figure}
    \centering
    \null \hfill \begin{subfigure}
        \centering
        \includegraphics[width=\kochsubfigwidth]{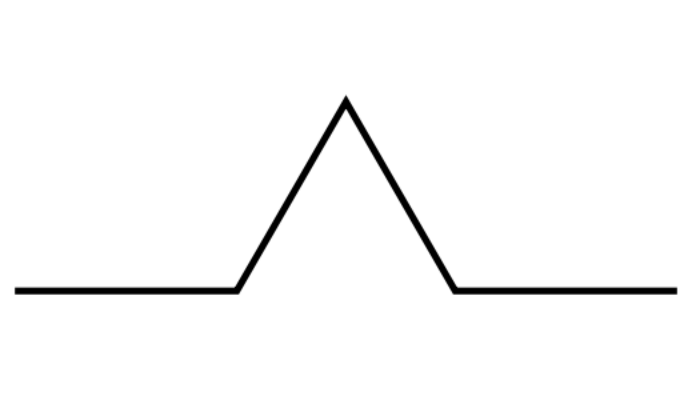}
    \end{subfigure}
    \hfil
    \begin{subfigure}
        \centering
        \includegraphics[width=\kochsubfigwidth]{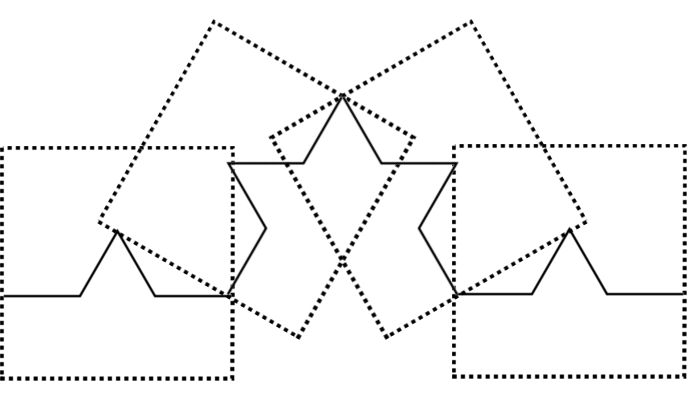} 
    \end{subfigure} \hfill \null
    \\
    \null \hfill \begin{subfigure}
        \centering
        \includegraphics[width=\kochsubfigwidth]{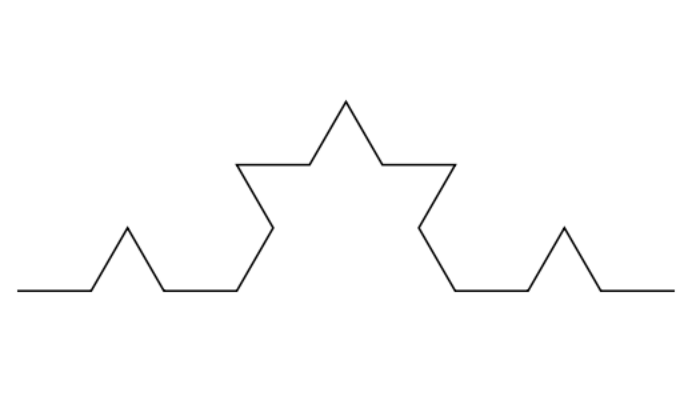} 
    \end{subfigure}
    \hfil
    \begin{subfigure}
        \centering
        \includegraphics[width=\kochsubfigwidth]{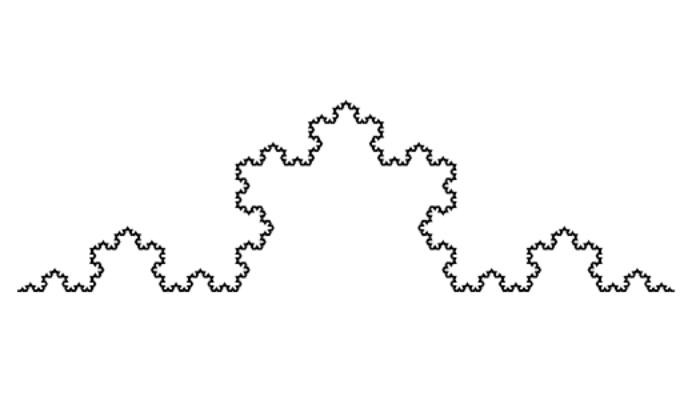} 
    \end{subfigure} \hfill \null
    \caption{From left to right: First row: The first iteration of the Koch IFS fractal; the second iteration, showing its composition as several scaled and rotated copies of the first iteration. Second row: As in the second image, but without bounding boxes; the attractor of this process. }
    \label{fig:koch}
\end{figure}

A related common construction for self-similar curves is to start with a base figure, and then recursively replace parts of the current figure with a transformed copy of the entire figure. For example, the Koch curve (see Figure \ref{fig:koch}) is generated by beginning with the line segment between $(0,0)$ and $(0, 1)$ and recursively applying the four affine transformations shown in Equation \ref{eqn:koch}. 

\begin{equation}
    \centering
    \begin{tabular}{l}
         $T_1 = \left[ \begin{array}{ccc}
              \frac{1}{3} & 0 & 0 \\
              0 & \frac{1}{3} & 0 \\
              0 & 0 & 1 \\
         \end{array} \right] $ 
         $ T_2 = \left[ \begin{array}{ccc}
              \frac{1}{6} & -\frac{\sqrt{3}}{6} & \frac{1}{3} \\
              \frac{\sqrt{3}}{6} & \frac{1}{6} & 0 \\
              0 & 0 & 1 \\
         \end{array} \right] $ \\
         $ T_3 = \left[ \begin{array}{ccc}
              \frac{1}{6} & \frac{\sqrt{3}}{6} & \frac{1}{3} \\
              -\frac{\sqrt{3}}{6} & \frac{1}{6} & 0 \\
              0 & 0 & 1 \\
         \end{array} \right] $ 
         $ T_4 = \left[ \begin{array}{ccc}
              \frac{1}{3} & 0 & \frac{2}{3} \\
              0 & \frac{1}{3} & 0 \\
              0 & 0 & 1 \\
         \end{array} \right] $ \\
    \end{tabular}
    \label{eqn:koch}
\end{equation}

\subsection*{IFS Inverse Problem}
Now that I have defined IFS fractals, I am ready to introduce the \textit{inverse IFS problem}. Simply put, this is the problem of finding a set of affine transformations having a specific image as their attractor. This problem is also called the ``fractal image compression'' problem, motivated by storing a set of affine transformations instead of the pixels of the original image. The inverse IFS problem is a classic problem, and a wide variety of approaches have been proposed. An early approach due to Barnsley et. al \cite{barnsley1986solution} proposed optimizing a loss function that measured distortion of the target region under the Hausdorff metric (``the Collage Theorem''). Other approaches have been proposed which utilize the convex hull of the target region \cite{hocevar2002inverse}; genetic algorithms \cite{lutton1995mixed}; wavelet transforms \cite{rinaldo1994inverse}; EM-like algorithms \cite{bloem2017expectation}; and in some cases, neural networks \cite{la2018iterated,graham2021applying}. However, no general solution is known to exist. Prior work by Melnik et. al \cite{melnik1998gradient}, similar to this paper, uses gradient descent to optimize IFS fractals. Those authors train a recurrent neural network $R(X)$ to iteratively transform a set of points $X$, with the Hausdorff metric between $X$ and $R(X)$ as the loss function. In contrast to this paper, Melnik et. al's work does not generate an explicit set of affine transformations. I stress here that I do not claim to have solved the inverse IFS problem in this paper; I am using this problem to demonstrate the utility of differentiating through fractal rasterization. 

\subsection*{Signed Distance Functions}
Signed distance functions are a way to represent the geometry of a region in Euclidean space. An SDF is the distance between a point in space and the boundary of a shape; points inside the shape are assigned the negative of this distance. SDFs have several properties that make them attractive in a rendering context: 1) for many geometric primitives exact SDF formulae are known \cite{quilez2999sdfs}; 2) in Euclidean space, the SDF of a region is differentiable almost everywhere; and 3) SDFs compose nicely with each other (for example, in Boolean operations) and with affine transformations. These properties (amongst others) have lead to widespread integration of SDFs into a variety of machine learning approaches in recent years \cite{park2019deepsdf,vicini2022differentiable,osher2003constructing, sitzmann2020metasdf, chan2023generative, chen2021learning}.

One specific characteristic of SDFs that will be necessary later in this paper is how a SDF behaves under affine transformation. Let $d_\mathcal{S}(x)$ be a SDF for a region $\mathcal{S}$, and let $T(x)$ be an affine transformation with scale parameter $s$. Then
\begin{equation}
    d_{T(\mathcal{S})}(x) = \frac{1}{s} d(T^{-1}(x)) 
    \label{eqn:sdf_affine}
\end{equation}
In other words, we can easily calculate a new SDF that represents an affinely transformed version of $\mathcal{S}$ by evaluating $d$ on $x$ but with the inverse transform of $T$. Note that this only holds when $T$ scales space uniformly - this equation does not hold for shear deformations. 
\subsection*{Differentiable Rendering}
Automatic Differentiation (``autodiff'' or AD) is a programming paradigm that has aided in the explosive growth of deep learning. The basic idea of autodiff is that all basic operations in the programming language are defined in a way that includes their derivative. This means that we can use the chain rule to compute the derivative of our entire program with respect to its inputs, enabling optimization via gradient descent. Recent work has successfully used autodiff to write full rasterization pipelines which are differentiable. Mihai et al. \cite{mihai2021differentiable} detail how to (differentiably) turn an implicit line segment into a pixelated picture of a line segment; in the next section I take the ability to do this as a given, and use it to produce fractal images. 

Some recent work has successfully used autodiff and signed distance functions to write full rasterization and rendering pipelines which are differentiable. The key insight is that we can render pixels by finding the distance between them and our geometric primitive $P$ \cite{mihai2021differentiable}. To render the pixel at location $(i,j)$, I take the exponential of the distance between the point $(i,j)$ and the primitive: $\text{pixel}_{i,j} = \exp(-d((i,j), P)^2 / \sigma^2)$. Because this pixel value function is a differentiable function of distance, and distance is a differentiable function of the parameters of $P$, I can tune these parameters using gradient descent. This allows fitting geometric primitives to images.

\section*{Rendering Process}
The components of my proposed differentiable fractal are 
\begin{itemize}
    \item \textbf{Control points:} a set of variables $\mathcal{P} = \{ p_1, p_2 \ldots p_n \}$, where all $p_i \in \mathbb{R}^2$. 
    \item \textbf{Symmetry pattern:} a sequence $L$ of pairs of subsets of $\mathcal{P}$: $L = \{ (P_1, R_1), (P_2, R_2), \ldots (P_m, R_m) \}$ with all $P_i, R_i \subset \mathcal{P}$. 
    \item \textbf{SDFs:} A set $\mathcal{D}$ of SDFs $d_1 \ldots d_l$, which may or may not also be defined in terms of the points $p_i$.
\end{itemize}
To make the above more specific, the Koch curve could be parametrized with a set of 5 control points $\mathcal{P} = \{p_0, p_1, p_2, p_3, p_4\}$. To get the same self-similar behavior as the Koch curve, we would take our symmetry pattern as:
\begin{multline*}
   L = \left\{ (\{p_0, p_4 \}, \{p_0, p_1 \}), (\{p_0, p_4 \},\{p_1, p_2 \}), \right. \\ \left. (\{p_0, p_4 \},\{p_2, p_3 \}), (\{p_0, p_4 \},\{p_3, p_4 \}) \right\} 
\end{multline*}

In other words, $L$ is specifying that the line segment $\{p_0, p_4 \}$ should be transformed into each of the line segments $\{p_i, p_{i+1} \}$ for $i = 0, 1, 2, 3$. Our SDFs $d_i$ would be functions representing the distance to each of these line segments (a closed formula for this distance is defined in terms of the endpoints of each segment). This choice of $(\mathcal{P},L,\mathcal{D})$ has the same replacement logic as the Koch curve, making it possible (assuming all intermediate operations are differentiable) to optimize the locations of the $p_i$ so that the linear transformations $T_k$ match those in Equation \ref{eqn:koch}.

\begin{figure}
    \centering
    \begin{subfigure}
        \centering
        \includegraphics[width=\fractalsubfigwidth]{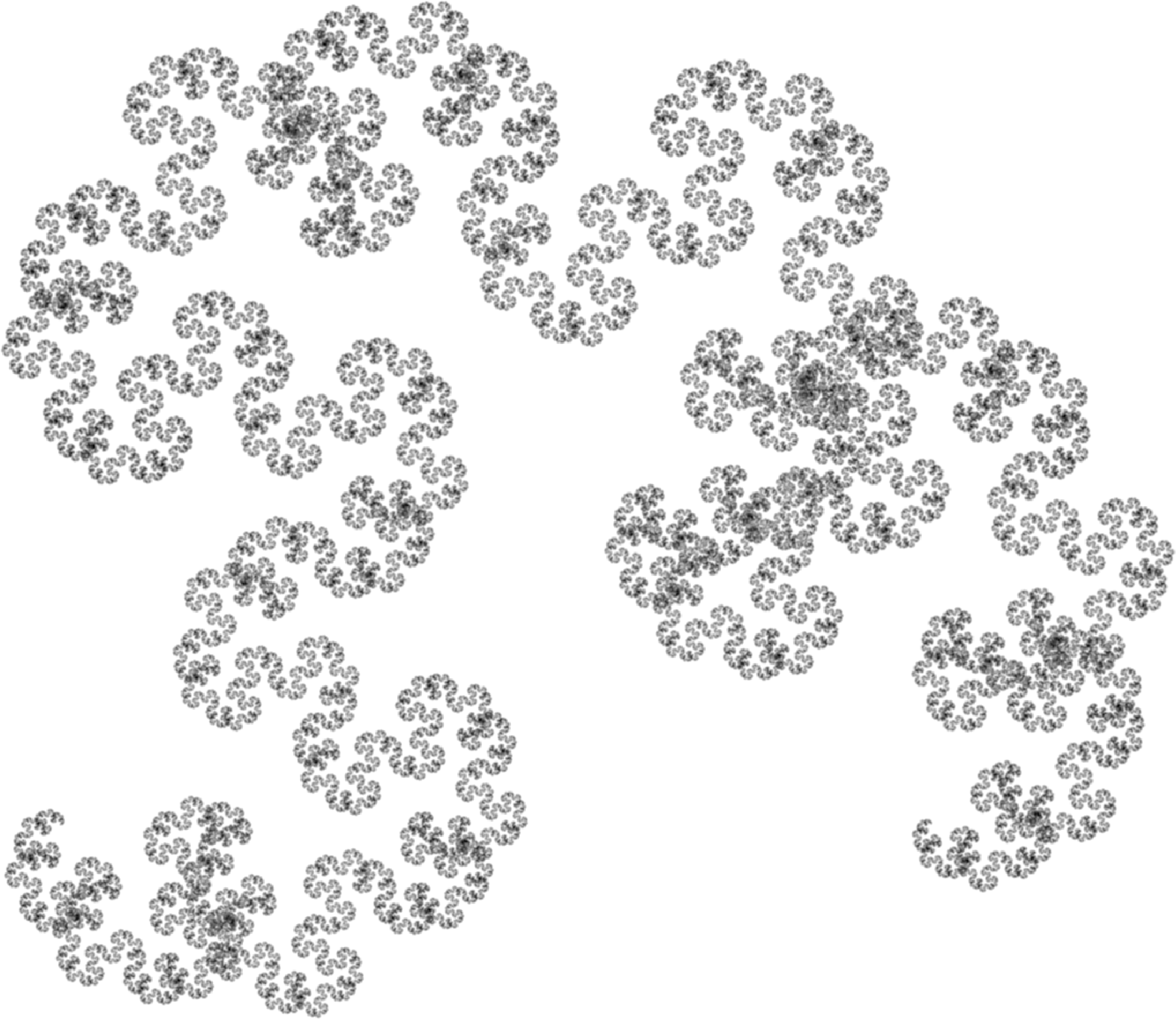}
    \end{subfigure}
    \hfill
    \begin{subfigure}
        \centering
        \includegraphics[width=\fractalsubfigwidth]{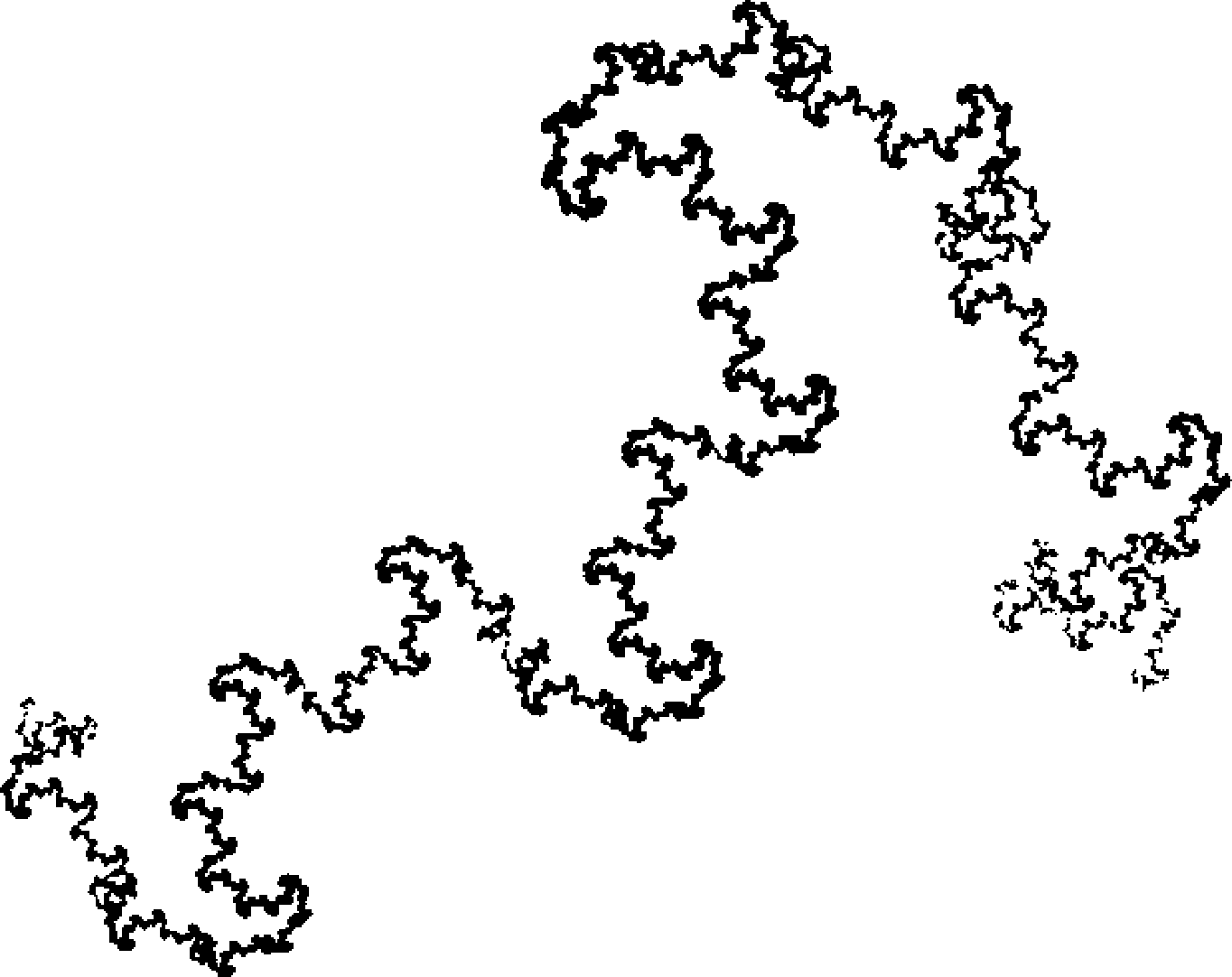} 
    \end{subfigure}
    \hfill
    \begin{subfigure}
        \centering
        \includegraphics[width=\fractalsubfigwidth]{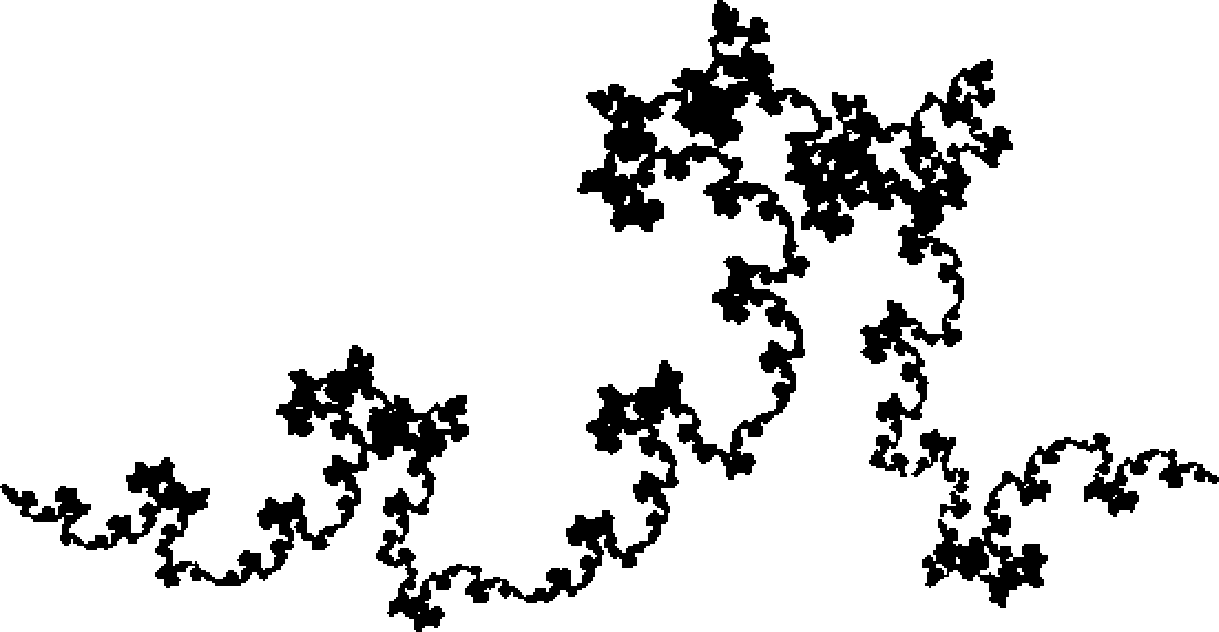} 
    \end{subfigure}
    \hfill
    \begin{subfigure}
        \centering
        \includegraphics[width=\fractalsubfigwidth]{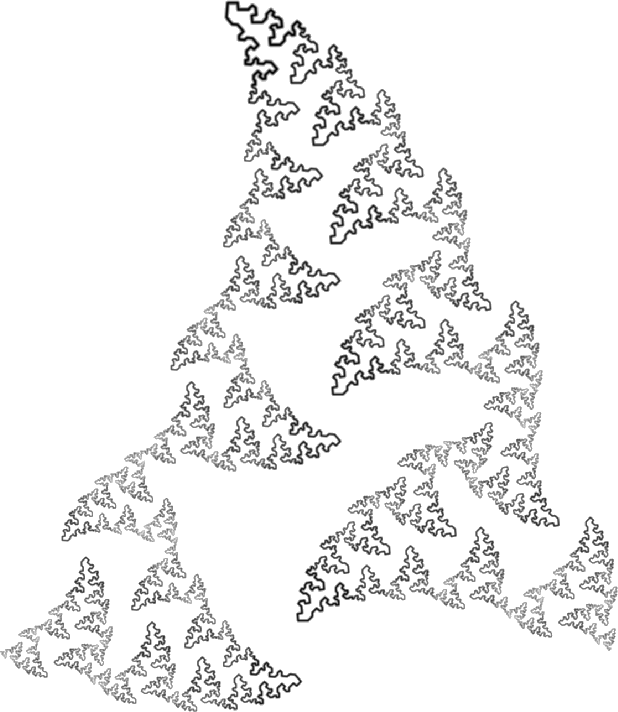} 
    \end{subfigure}
    \hfill
    \begin{subfigure}
        \centering
        \includegraphics[width=\fractalsubfigwidth]{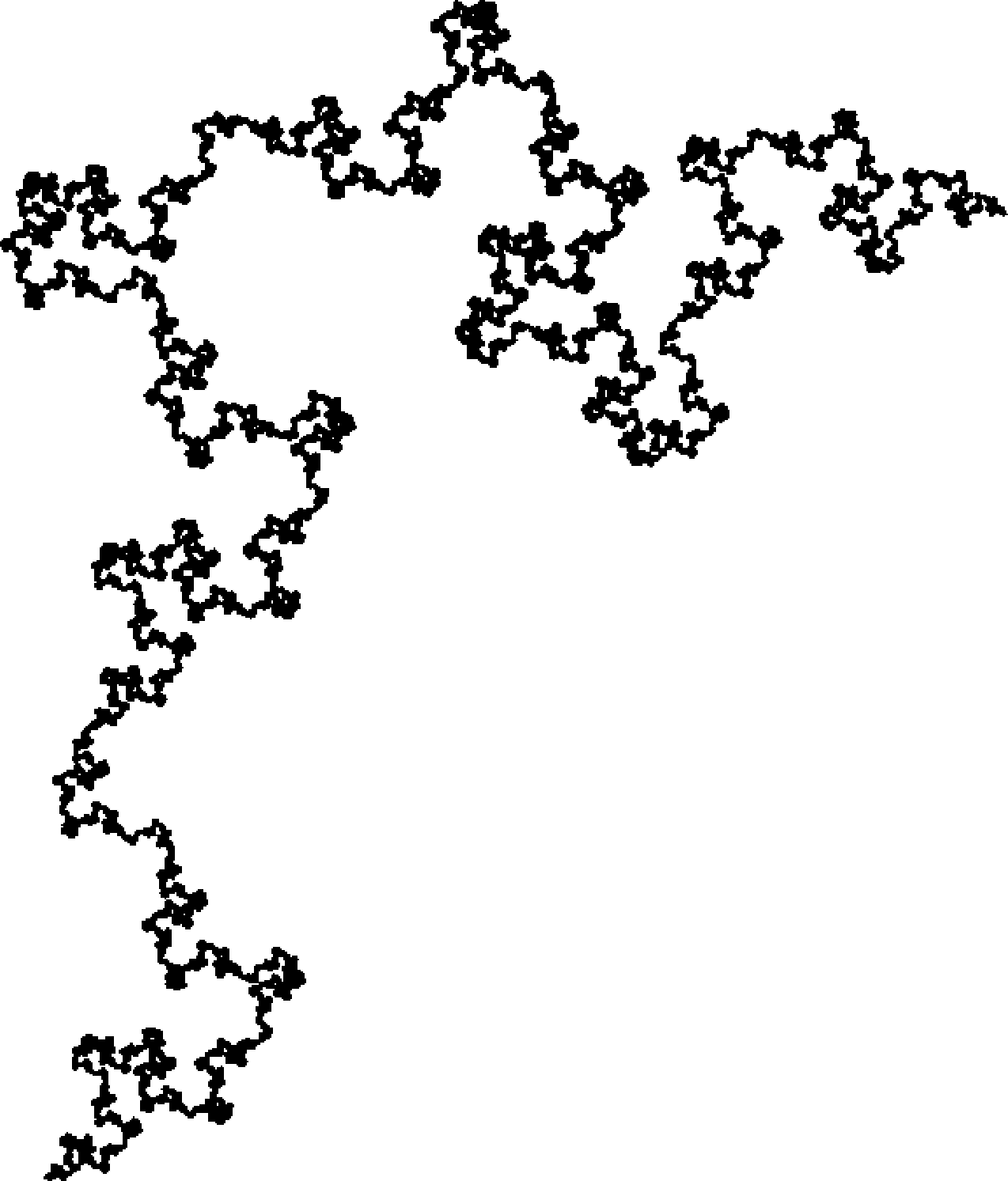} 
    \end{subfigure}

    \vspace{.001cm}
    \centering
    \begin{subfigure}
        \centering
        \includegraphics[width=\fractalsubfigwidth]{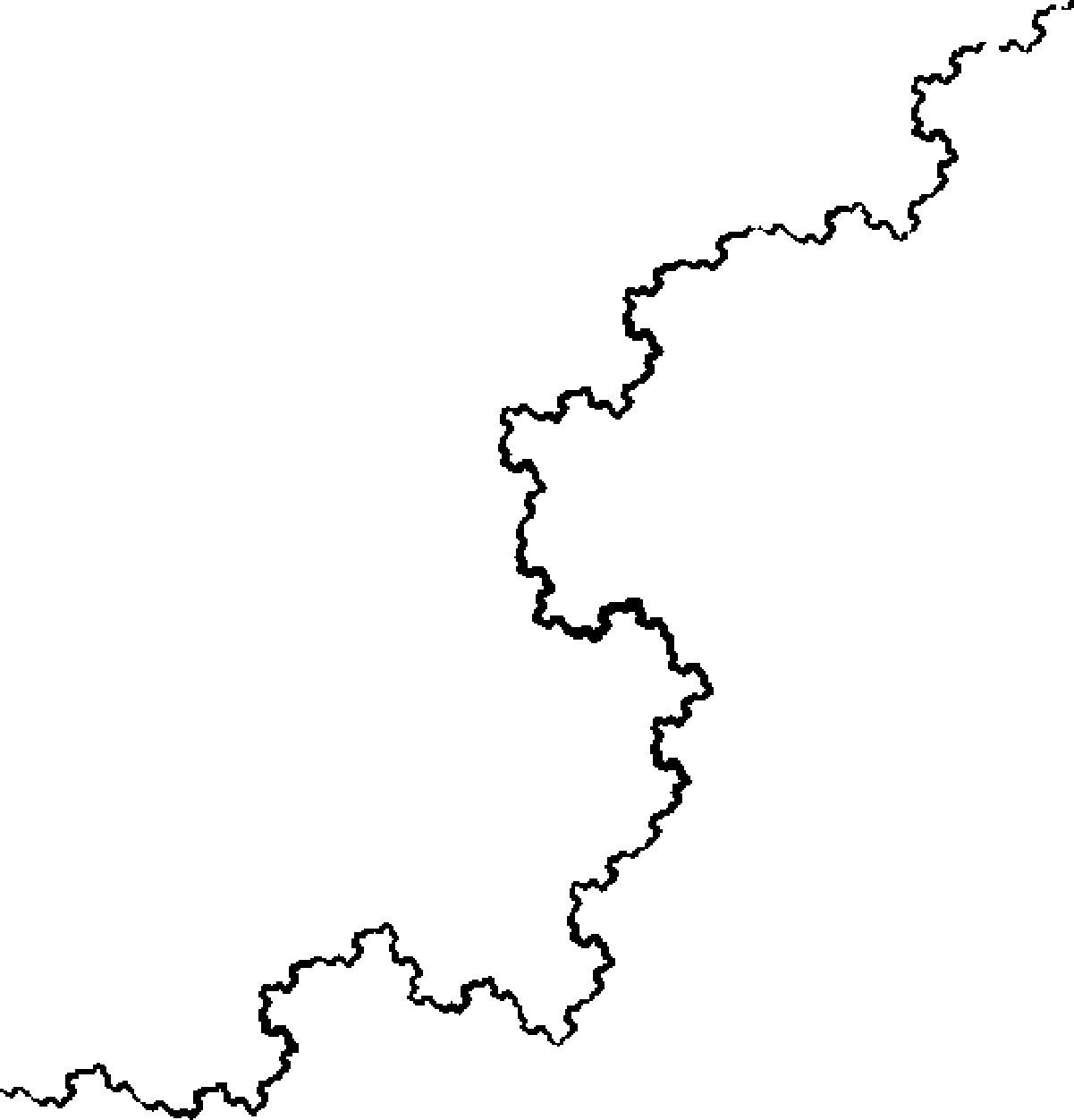}
    \end{subfigure}
    \hfill
    \begin{subfigure}
        \centering
        \includegraphics[width=\fractalsubfigwidth]{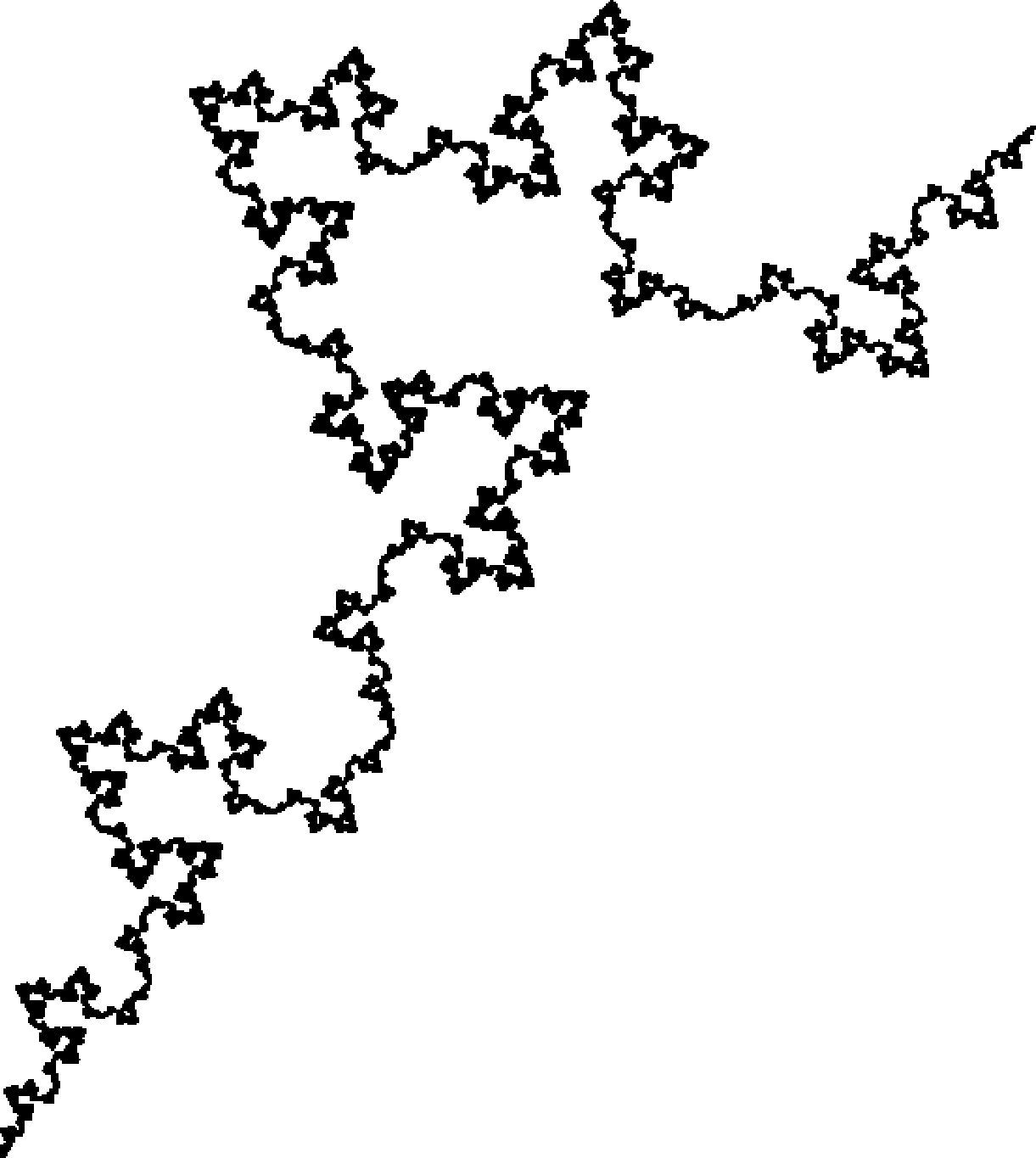} 
    \end{subfigure}
    \hfill
    \begin{subfigure}
        \centering
        \includegraphics[width=\fractalsubfigwidth]{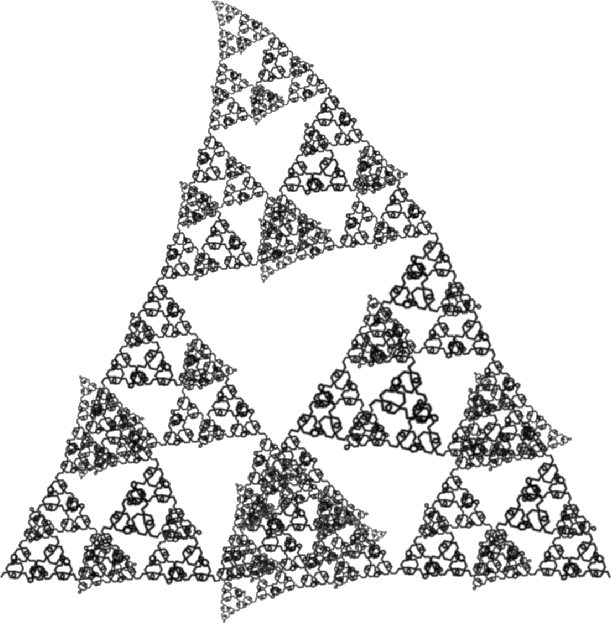} 
    \end{subfigure}
    \hfill
    \begin{subfigure}
        \centering
        \includegraphics[width=\fractalsubfigwidth]{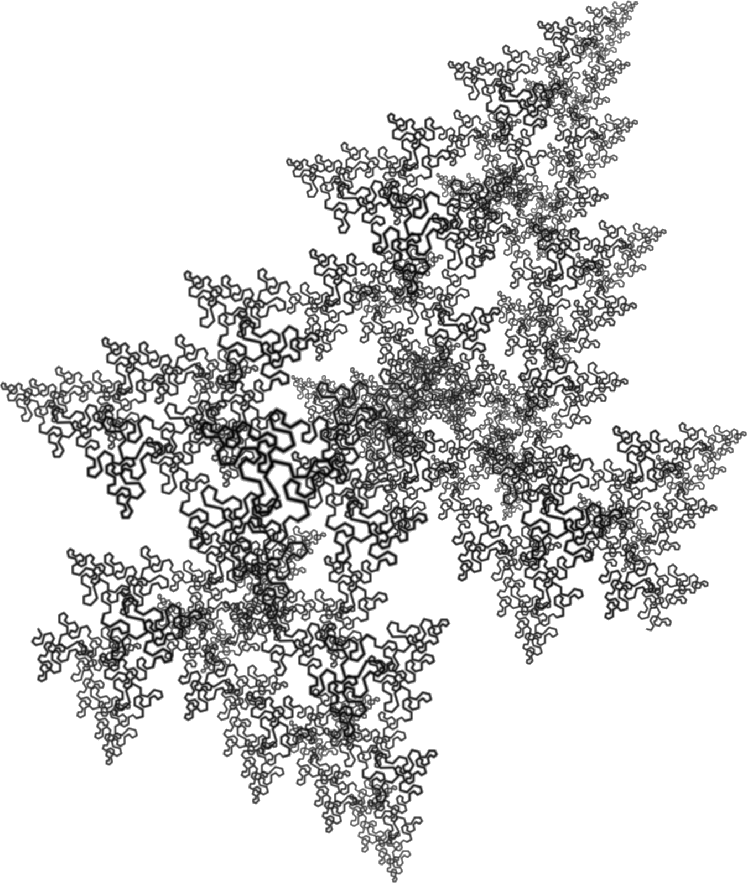} 
    \end{subfigure}
    \hfill
    \begin{subfigure}
        \centering
        \includegraphics[width=\fractalsubfigwidth]{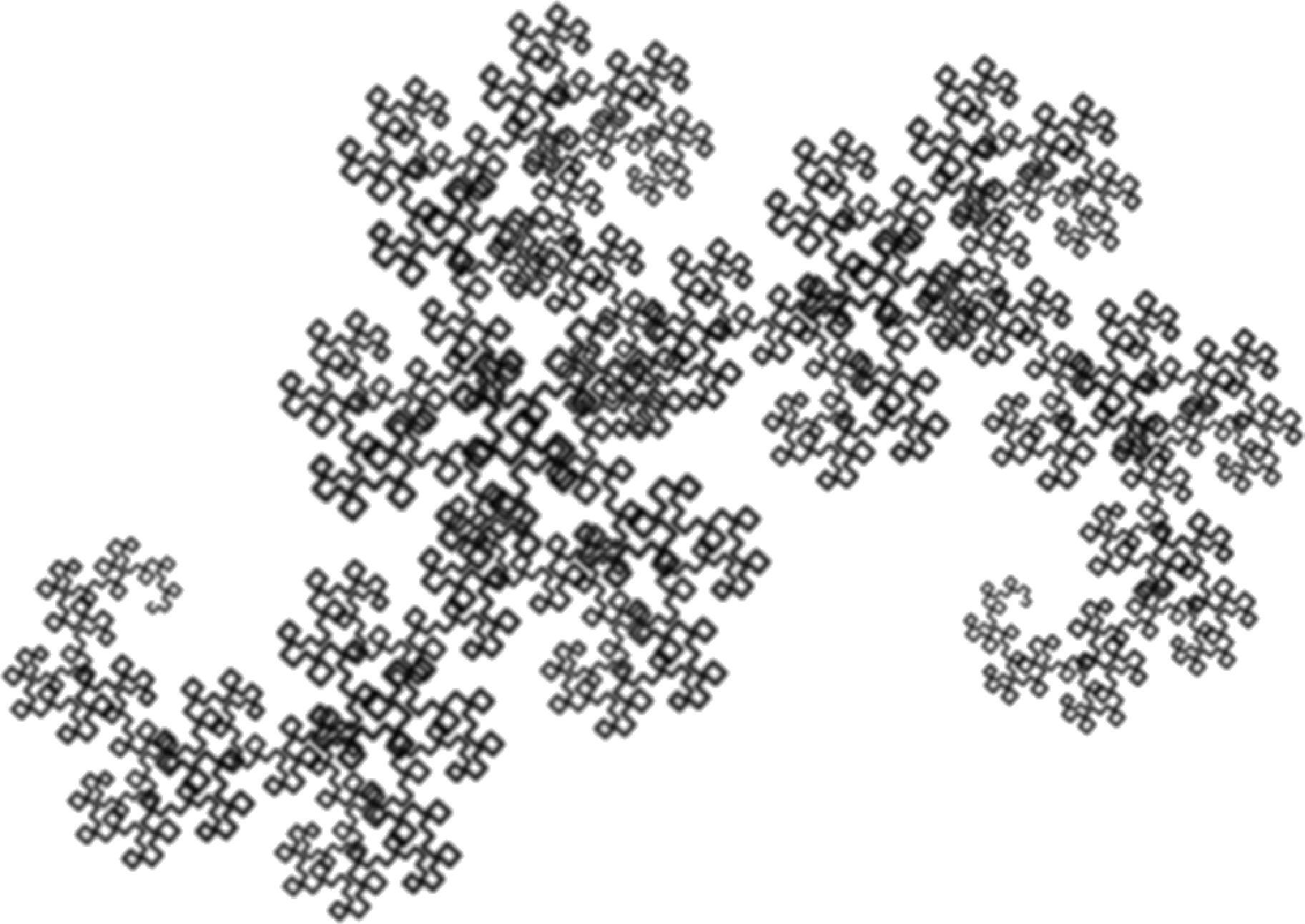} 
    \end{subfigure}

    \caption{Additional example fractals generated with the method outlined in this paper.}
    \label{fig:example_img}
\end{figure}

Once defined, the basic process in generating a fractal figure from the pieces described above is as follows:
\begin{enumerate}
    \item For each $i$ in $1 \ldots m$, find the linear transformation $T_i$ which (possibly approximately) maps $P_i$ to $R_i$.
    \item Up to a set maximum number of recursions $K$, replace the SDFs in $\mathcal{D}$ with the result of applying each transformation $T_i$ to all of the current elements of $\mathcal{D}$.
    \item Compute the distance $d((i,j))$ from each pixel $(i,j)$ to each SDF $d \in \mathcal{D}$.
    \item Color the pixel $(i,j)$ with the value $\exp( - (\min_{d \in \mathcal{D}} d((i,j)))^2 / \sigma^2)$. $\sigma$ is a scale parameter that determines the ratio of pixel coordinates to the coordinates of the $p_i$.
    \item (If optimizing) compare to target and compute loss.
\end{enumerate}
To get an IFS fractal image that looks like our source image, I implemented the above steps in Pytorch. The position of the control points was tuned with Adam \cite{kingma2014adam} to minimize the loss, i.e. so that our rendered image matches some target image. The maximum recursion depth, $K$, was set to be as deep as possible within GPU memory constraints. I will now discuss some of the details of the rendering pipeline; readers who want to recreate the images in this paper should refer to the Github repository. Figure \ref{fig:example_img} demonstrates a variety of images with the same symmetry set as the Koch curve but different initial conditions (and trained against a different target image). Figure \ref{fig:all_fract_exs} demonstrates the result of optimizing with different symmetry patterns, such as the replacement rule that yields the Sierpi\'{n}ski carpet. Figure \ref{fig:all_fract_exs} also includes the result of applying this optimization procedure to images that have the ``wrong" symmetry patter; that is, where $L$ does not actually match any symmetries in the target image. It may be possible to learn these symmetry patterns automatically; see Future Work. 

\subsection*{Implementation Details}
In this section I describe several of the specific design choices I made in my fractal renderer. 
\paragraph{Initial Conditions.}
As described, this system is extremely sensitive to initial conditions, which limits its applicability. For the Koch curve, I was able to consistently get optimization to converge by initializing the control points $p_i$ along a best-fit line of the black pixels in the target image.

\paragraph{Calculating Transformations.}
An important ingredient in the above process is the calculation of the linear transformation which takes our endpoints to each line segment. To do this for two line segments $(e_1, e_2)$ and $(p_1, p_2)$, we: a) find the translation that takes $e_1$ to $p_1$, b) find the rotation that aligns the two vectors, and then c) scale $(e_1, e_2)$ to have the same length as $(p_1, p_2)$. The composition of these three linear transformations is the one we want. 

\begin{figure}
    \centering
    \includegraphics[width=.68\linewidth]{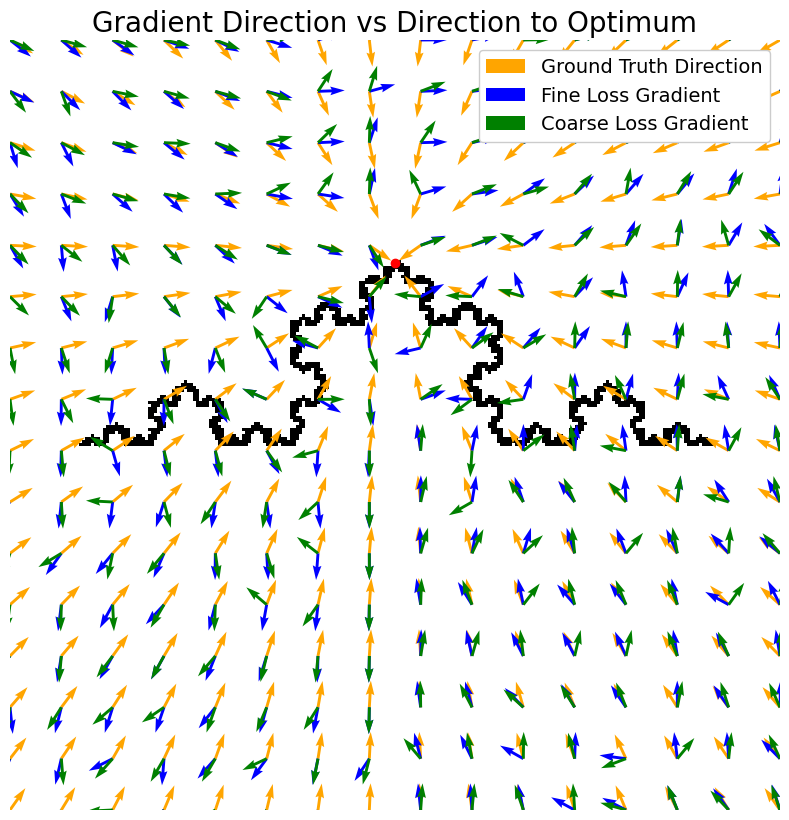}
    \caption{Vector fields illustrating the gradient of error as $p_2$ is varied in the Koch curve construction, while all other points are held constant at their optimal positions: a) always pointing toward the optimal location (orange); b) the gradient of the fine-scale loss only; and c) the gradient of the multiscale loss. While both gradient vector fields are divergent, the multiscale loss is slightly more aligned with the always-optimal field.}
    \label{fig:vecfield}
\end{figure}

\paragraph{Loss Function.}
Following the example of \cite{mihai2021differentiable}, I used Multiscale Mean Squared Error (MMSE) as the loss function. MMSE is identical to mean-squared error, but is summed over multiple pooled copies of the image. I experimented with two variants of loss function: computing MMSE over the pixel values in rasterized images, versus the MMSE between the raw distance values calculated by the SDF. Both of these loss functions worked reasonably well, but tended to get stuck in local optima (see Figure \ref{fig:koch_minima} for examples). Figure \ref{fig:vecfield} illustrates why this might be happening: even arbitrary close to the boundary of the fractal figure, the gradient of the loss (in this case, pixel loss) can point away from the location of the true optimum. Note in this image that the green arrows, representing MMSE, point toward the true optimum slightly more often than MSE at only the finest scale (in blue). 

\newcommand{\fracfigwid}{.29\linewidth}

\begin{figure}
    \centering
    \begin{tabular}{ccc}
        Target Image & Initial State & Converged State \\
        \includegraphics[width=\fracfigwid]{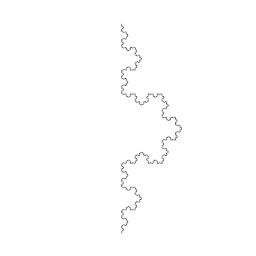} &
        \includegraphics[width=\fracfigwid]{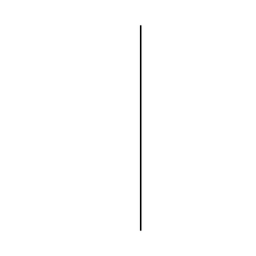} &
        \includegraphics[width=\fracfigwid]{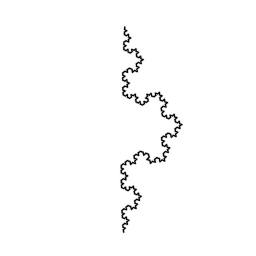} \\
        \includegraphics[width=\fracfigwid]{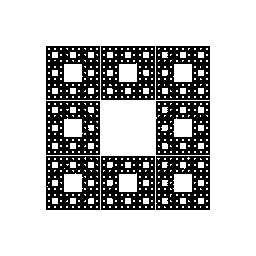} &
        \includegraphics[width=\fracfigwid]{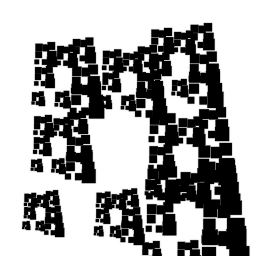} &
        \includegraphics[width=\fracfigwid]{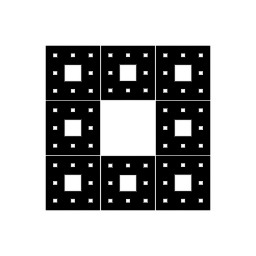} \\
        \includegraphics[width=\fracfigwid]{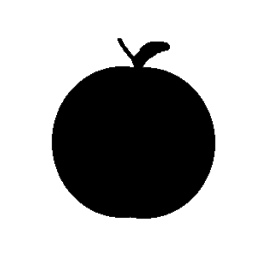} &
        \includegraphics[width=\fracfigwid]{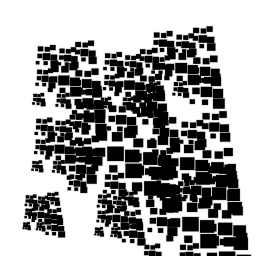} &
        \includegraphics[width=\fracfigwid]{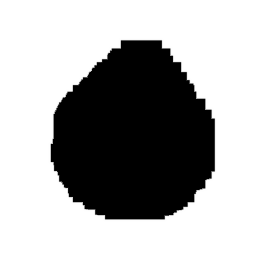} \\
        \includegraphics[width=\fracfigwid]{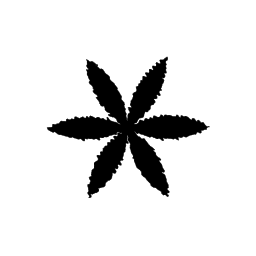} &
        \includegraphics[width=\fracfigwid]{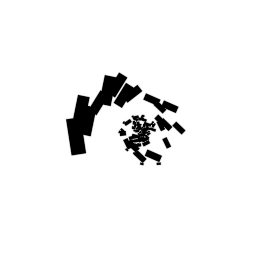} &
        \includegraphics[width=\fracfigwid]{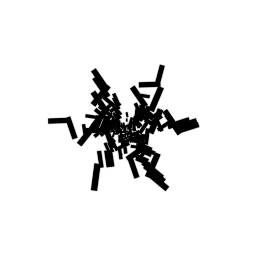} 
    \end{tabular}
    \caption{Multiple examples of learning IFS attractors. In each row from top to bottom: the Koch curve; the Sierpi\'{n}ski carpet; an apple from the MPEG-7 shape dataset; a flower from the MPEG-7 shape dataset. }
    \label{fig:all_fract_exs}
\end{figure}

\newcommand{\localminwidth}{.19\linewidth}
\begin{figure}
    \centering
    \includegraphics[width=\localminwidth]{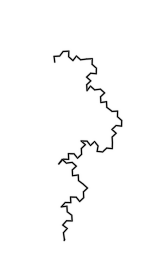} \hfill
    \includegraphics[width=\localminwidth]{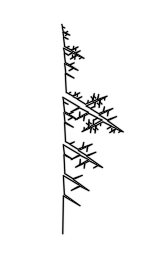} \hfill
    \includegraphics[width=\localminwidth]{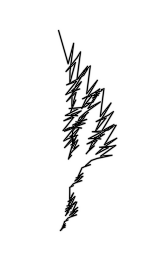} \hfill
    \includegraphics[width=\localminwidth]{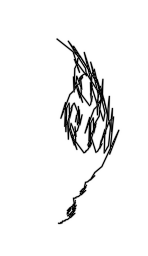} \hfill
    \caption{Examples of local optima encountered while optimizing a set of affine transformations to fit the Koch curve.}
    \label{fig:koch_minima}
\end{figure}

\paragraph{Code Repository}
All code for the operations described in this paper is available at \repourl.

\section*{Conclusion and Future Work}
This paper presents initial evidence that it is possible to find the parameters of IFS fractals and self-similar curves using automatic differentiation and gradient descent. However, there are many open questions that still need to be addressed in order to make this a viable technique for solving the IFS inverse problem. In Figure \ref{fig:all_fract_exs}, the Sierpinski and Koch examples both use the known, correct symmetry pattern for these IFSs. This is somewhat unfair, since learning the symmetry pattern in a domain is a harder problem than finding an affine transformation between point sets. Initial attempts at learning a symmetry pattern as a weighted combination of control points proved unsuccessful (optimization diverged in every case). One of the other IFS solving approaches may help here, as might recent machine learning work in automated discovery of symmetry in geometric objects \cite{shi2020symmetrynet}.

Additionally, it is currently unclear what determines whether optimization will converge, diverge, or converge to a local minimum. Very careful initialization of control point locations is necessary to get results that resemble the target image. Developing a procedure for appropriately initializing control points given a target image is necessary to apply this technique to arbitrary images. The examples in Figure \ref{fig:koch_minima} seem to indicate that local minima are related to times when the figure crosses itself; a barrier function that prevents self-crossings could keep the optimization from getting trapped in these local optima. 

Finally, there are also several modifications that could be make to make the process more efficient, such as the rendering tricks mentioned in \cite{hepting1991rendering}. One major area for future work is investigating the viability of coarse-to-fine optimization procedures, which are a mainstay of many graphics algorithms and optimization procedures. 

Overall, this paper represents a proof-of-concept of differentiating through fractal rasterization. More work is necessary to determine the bounds and capabilities of the proposed approach.

\printbibliography

\end{document}